\documentstyle[sprocl,psfig]{article}

\def\Journal#1#2#3#4{{#1} {\bf #2}, #3 (#4)}

\def\PLB{{\em Phys. Lett.}  B}
\def\PRL{\em Phys. Rev. Lett.}
\def\PRD{{\em Phys. Rev.} D}

\def\be{\begin{equation}}
\def\ee{\end{equation}}
\def\bea{\begin{eqnarray}}
\def\eea{\end{eqnarray}}

\begin{document}

\title{
CURRENT ISSUES IN QUARKONIUM PRODUCTION\footnote{
Talk presented at the DIS 2000, Liverpool, England, April 25-30, 2000.
}
}

\author{JUNGIL LEE}

\address{II. Institute of Theoretical Physics,\\
University of Hamburg, 22761 Hamburg, Germany
}

\maketitle\abstracts{
As a stringent test of the nonrelativistic QCD factorization approach
for inclusive heavy-quarkonium production in high-energy hadron collision,
the polarized heavy-quarkonium production at the Tevatron is discussed.
The polarization of prompt $J/\psi$ at the Tevatron is predicted
within the nonrelativistic QCD factorization framework.
The contribution from radiative decays of P-wave charmonium states decreases,
but does not eliminate, the transverse polarization at large transverse 
momentum.  The prediction agrees with measurements
from the CDF collaboration at intermediate values of $p_T$,
but disagrees at the large values of $p_T$ measured.
Recent preliminary data of heavy quarkonia from
CDF and theoretical/experimental prospects 
related to the Run II heavy quarkonium physics are also discussed.
}
As an effective field theory of QCD, the nonrelativistic QCD (NRQCD) 
factorization approach~\cite{B-B-L} has had a great success in describing 
the inclusive production and decay of heavy quarkonium systems 
which could not be explained by a conventional phenomenological model, 
the so-called color-singlet model (CSM). 
The color-octet mechanism of NRQCD, which is absent in the CSM, 
presented a resonable solution~\cite{B-F} to the anomalous production rate of 
$\psi'$ and $J/\psi$  at the Fermilab Tevatron~\cite{CDF-psi}.
In high-energy $p \bar p$ collisions,
the dominant contribution to the charmonium production rate at large $p_T$
comes from gluon {\it fragmentation}~\cite{B-Y}.
At leading order in $\alpha_s$, a $Q \overline{Q}$ pair with small
relative momentum created by the virtual gluon is in a color-octet
$^3S_1$ state with the same transverse polarization as the almost
on-shell gluon.
The enhancement of the production cross section due to the gluon propagator
makes the fragmentation process dominate.
Since the short-distance contribution has two powers of $\alpha_s$ less than
the CSM gluon fragmentation channel, the high-$p_T$ S-wave charmonium
production rate should be dominated by the process from the transition 
of color-octet $^3S_1$ state even though there is a suppression factor 
$v^4$ coming from a  double chromoelectric dipole transition 
in the non-perturbative process~\cite{B-F}.

The NRQCD factorization approach to inclusive quarkonium 
production~\cite{B-B-L} also 
makes the remarkable prediction that in hadron collisions
$J^{PC} = 1^{--}$ quarkonium states
should be transversely polarized at large $p_T$~\cite{Cho-Wise}.  
Due to the heavy-quark spin symmetry of the NRQCD interaction Lagrangian 
governing the chromoelectric dipole transition and
the fact that the transition of the $Q \overline{Q}$ pair into a 
physcal $S-$wave charmonium is dominated by a double chromoelectric
dipole transition, 
the polarizarion of $J/\psi$ mesons produced at sufficiently large $p_T$ 
should be transversely polarized.
Recent measurements at the Tevatron by the CDF collaboration seem to be in 
contradiction with this prediction~\cite{CDF-psipol}.

A convenient measure of the polarization is 
$\alpha = (T-2L)/(T+2L)$, where $T$ and $L$ are the transverse
and longitudinal polarization fractions in the hadron CM frame.
The variable $\alpha$ 
describes the angular distribution of leptons from the decay
of the $J/\psi$ with respect to the $J/\psi$ momentum.
Beneke and Rothstein studied the dominant fragmentation mechanisms for
producing longitudinally polarized $1^{--}$ states~\cite{Beneke-Rothstein}.
For charmonium production at the Tevatron, 
one should also take into account the {\it fusion} 
contributions from parton processes $i j \to c \bar c + k$
since fragmentation does not yet dominate for most of the $p_T$ region.
The polarization variable $\alpha$
for direct $J/\psi$ and direct $\psi'$ at the Tevatron
have been predicted by Beneke and
Kr\"amer~\cite{Beneke-Kramer} and by Leibovich~\cite{Leibovich}.
They predicted that $\alpha$ should be small for $p_T$ below about 5 GeV,
but then should begin to rise dramatically.
On the contrary, the CDF data shows no sign of transverse polarization 
of direct  $\psi'$
at large $p_T$~\cite{CDF-psipol}.

The CDF collaboration has also measured the
polarization of {\it prompt} $J/\psi$~\cite{CDF-psipol} 
({\it i.e.} $J/\psi$'s that do not come from the decay of $B$ hadrons).
The advantage of this measurement is that the number of $J/\psi$ events
is larger than for $\psi'$ by a factor of about 100.
On the other hand, theoretical predictions of the polarization of 
prompt $J/\psi$ are complicated since the prompt signal includes
$J/\psi$'s that come from decays of the higher charmonium
states $\chi_{c1}$, $\chi_{c2}$, and $\psi'$~\cite{CDF-chi,CDF-chi12}.
The polarization of $J/\psi$ from $\psi'$ not via $\chi_c$ is
straightforward to calculate, since the spin is unchanged by the transition.
The polarization of $J/\psi$ from $\chi_c$ and of
$J/\psi$ from $\psi'$ via $\chi_c$ is more complicated,
because the $\chi_{cJ}$'s are produced in various spin states
and they decay into $J/\psi$ through radiative transitions~\cite{BKL}.
Therefore, it is interesting to investigate the cascade effect in this problem.

The {\it NRQCD factorization formula} for the differential cross section for
the inclusive production of a charmonium state $H$ of momentum $P$
and spin quantum number $\lambda$ has the schematic form
\begin{equation}
d \sigma^{H_\lambda(P)} \;=\;
d \sigma^{c \bar c_n(P)} \;
        \langle O^{H_\lambda(P)}_n \rangle,
\label{sig-fact}
\end{equation}
where the implied sum on $n$ extends over
all the color and angular momentum states of the $c\bar c$ pair.
The $c \bar c$ cross sections $d \sigma^{c \bar c_n}$, which are independent of
$H$, can be calculated using perturbative QCD.
All dependence on the state $H$ is contained within the
nonperturbative NRQCD matrix elements(ME's) 
$\langle O^{H_\lambda(P)}_n \rangle$.
The Lorentz indices, which are suppressed in (\ref{sig-fact}),
are contracted with those of $d \sigma^{c \bar c_n}$
to give a scalar cross section.
The symmetries of NRQCD can be used to reduce
the tensor ME's $\langle O^{H_\lambda(P)}_n \rangle$
to scalar ME's $\langle O^H_n \rangle$ that are
independent of $P$ and $\lambda$.
One may calculate the  cross section for polarized quarkonium
once the relevant scalar ME's are known.

In $p \bar p$ collisions,  the parton processes that dominate the
$c \bar c$ cross section depend on $p_T$.
If $p_T$ is of order $m_c$, the {\it fusion} processes dominate.
These include the parton processes $i j \to c \bar c + k$,
with $i,j=g,q,\bar q$ and $q = u,d,s$.
At $p_T$ much larger than $m_c$,
the parton cross sections are dominated by
{\it fragmentation} processes.
To get a consistent LO evaluation  of polarization parameter $\alpha$
in this region, one should take care.
Since we are interested in the LO prediction of $\alpha$, we  should first
get a LO prediction of the longitudinal fraction $L/(T+L)$.
The LO contributions to $L$ are order-$\alpha_s^2$
processes $g\to c\overline{c}+g$, while the only
contribution to $T$ is the order-$\alpha_s$ process  
$g \to c \bar c_8(^3S_1)$ \cite{Beneke-Rothstein}.
Inclusion of the longitudinal part in the fragmentation process is 
crucial in the calculation of $\alpha$ even though it has only a small
contribution to the total cross section.
The fragmentation function is evolved using the
standard homogeneous timelike evolution equation.
Since $c \bar c_8(^3S_1)\to \psi_\lambda(nS)$ is included in both
the fragmentation and the fusion process, we interpolate
between the fusion cross section at low $p_T$ and the fragmentation
cross section at high $p_T$~\cite{BKL,Cho-Leibovich}. 
In all the other $c \bar c\to H$ channels, we use the fusion cross section.

The color-singlet ME's can be determined
phenomenologically from the decay rates for
$\psi(nS)\to \ell^+\ell^-$ and $\chi_{c2}\to\gamma\gamma$~\cite{Maltoni}.
The color-octet ME's are phenomenological
parameters and they are determined from production data at the 
Tevatron~\cite{CDF-psi,CDF-chi}.
In case of $\chi_{c}$,
we have a good agreement with the CDF data on the ratio 
$\chi_{c1}/\chi_{c2}$~\cite{CDF-chi12}.
More details on the analysis methods are explained in 
Refs.~\cite{BKL,Kniehl-Kramer}.
There are numerous theoretical uncertainties in the polarization 
calculation. In our analysis, we allow the variations in
PDF (CTEQ5L and MRST98LO)~\cite{PDF}, 
factorization and fragmentation scales ($\mu_T/2- 2\mu_T$),
charm quark mass ($1.45-1.55$ GeV), respectively,
where $\mu_T = (4 m_c^2 + p_T^2)^{1/2}$.
We also take into account the uncertainties in the ME's,
$\langle O_8(^1S_0) \rangle$ and 
$\langle O_8(^3P_0) \rangle$~\cite{BKL}.

\begin{figure}
\begin{center}
\begin{tabular}{c}
\psfig{file=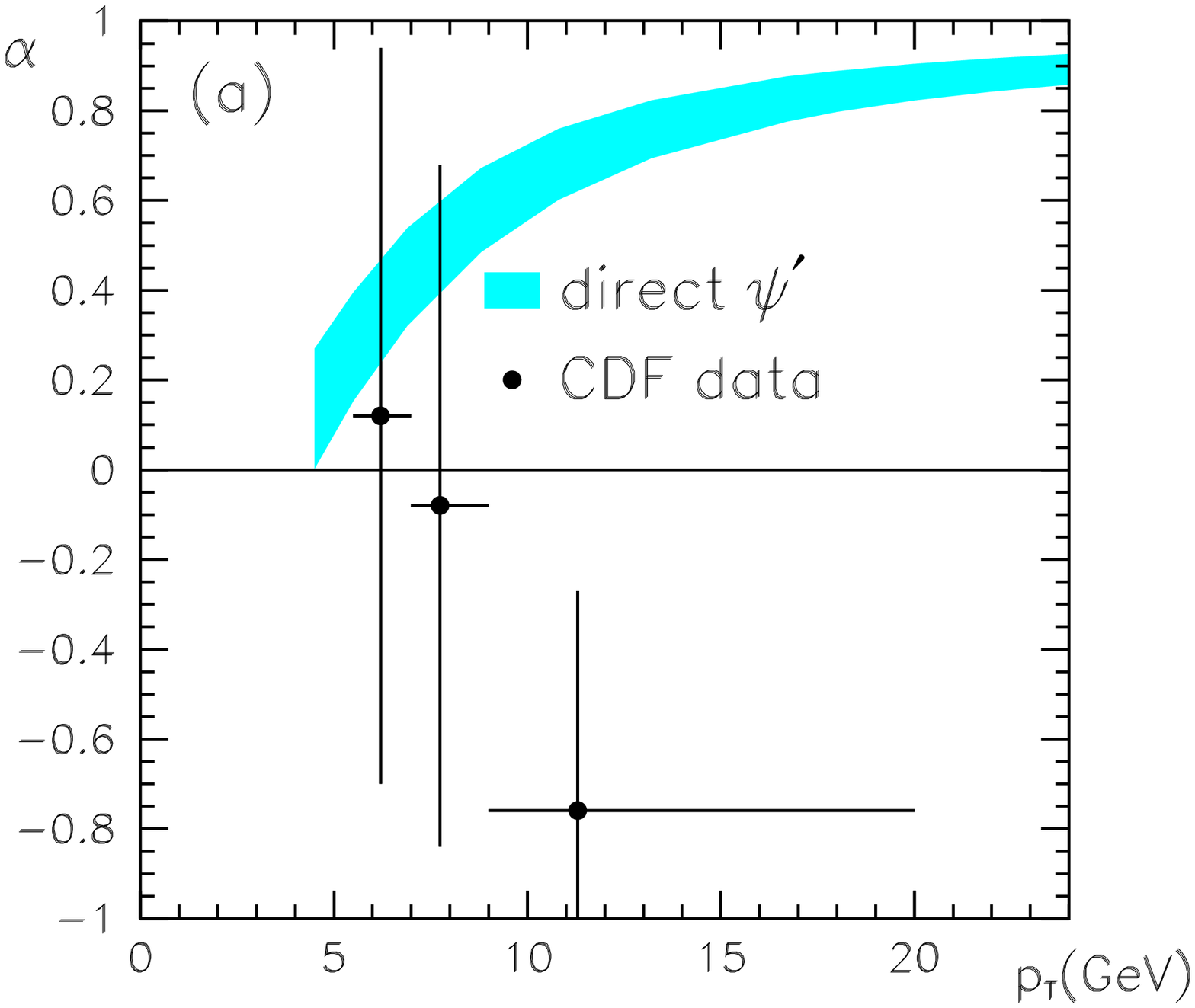,height=7.8cm}\\
\psfig{file=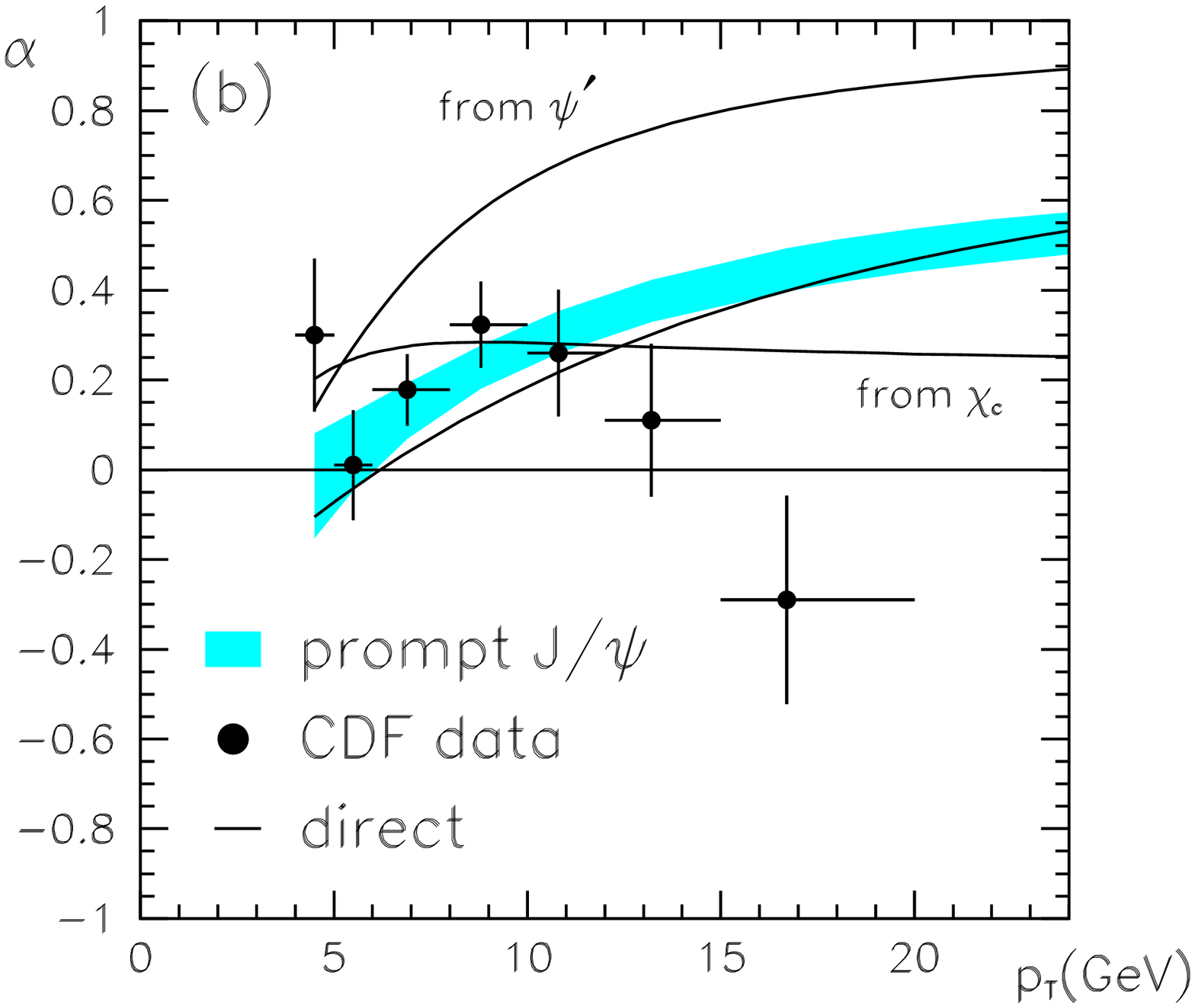,height=7.8cm}
\end{tabular}
\end{center}
\caption{
Polarization variable $\alpha$ vs. $p_T$
for (a) direct $\psi'$ and (b) prompt $J/\psi$
compared to CDF data.
}
\end{figure}

Let us compare our results for $\alpha$ with the CDF data.
We present our result in the form of an error band obtained by
combining in quadrature all the theoretical errors described previously.
In Fig.~1(a), we compare our result for direct $\psi'$
as a function of $p_T$ with the CDF data~\cite{CDF-psipol}.
But the error bars in the CDF data are too large to draw any
definitive conclusions.
We next consider the polarization variable $\alpha$ for prompt $J/\psi$.
The method for calculating cascade effect in prompt $J/\psi$ polarization 
is explained in Ref.~\cite{BKL}.
In Fig.~1(b), we compare our result for $\alpha$
as a function of $p_T$ with the CDF data~\cite{CDF-psipol}.
Our result for $\alpha$ is small around $p_T=5$ GeV, but it increases
with $p_T$.
Our result is in good agreement with the CDF measurement
at intermediate values of $p_T$,
but it disagrees 
in the highest $p_T$ bin, where the CDF measurement is consistent with 0.
The solid lines in Fig.~1(b) are the central curves of $\alpha$ for
direct $J/\psi$ and for $J/\psi$ from $\chi_c$. 
The $\alpha$ for direct $J/\psi$ is smaller than that
for direct $\psi'$, because the ME's are different~\cite{BKL}.
In the moderate $p_T$ region,
the contributions from $\psi'$ and from $\chi_c$ add to give an increase
in the transverse polarization of prompt $J/\psi$
compared to direct $J/\psi$.  In the high $p_T$ region,
the contributions from $\psi'$ and $\chi_c$ tend to cancel.

The CDF measurement of the polarization of prompt $J/\psi$
presents a serious challenge to the NRQCD factorization formalism
for inclusive quarkonium production.
The qualitative prediction that $\alpha$ should increase
at large $p_T$ seems inescapable.
However, it is still worthwhile to investigate the NLO effect 
in the fragmentation processes, since the dramatic discrepancy appears
at large $p_T$. Recently, Braaten and Lee performed the full NLO 
calculation of the color-octet $^3S_1$ gluon fragmentation function 
for polarized heavy quarkonium~\cite{BL}.
It is not sufficient to get NLO prediction of $\alpha$ for prompt $J/\psi$, 
but one can first study the production rate at large $p_T$ in NLO accuracy.
Full NLO calculation of fusion process in NRQCD framework is also necessary 
in order to get more precise values for the ME's.

In Run II of the Tevatron, the data sample of heavy quarkonium should be
more than an order of magnitude larger than in Run I.
This will allow for both the production rate and the polarization 
to be measured with higher precision and out to larger values of $p_T$.
The progress of the  experimental analysis in Run I will advance further. 
The NRQCD ME's for both the charmonium and bottomonium systems will 
be determined with higher accuracy. 
Especially, the accuracy of the color-octet $^3S_1$ ME for the $\psi'$ 
meson will be greatly increased since one can probe the high-$p_T$ region 
where fragmentation dominates and which currently has little statistics.
The measurements of the polarization parameter $\alpha$ 
for prompt $J/\psi$ and $\psi'$ will be more precise. 
It will also be possible to measure $\alpha$'s for
direct $J/\psi$ and  $J/\psi$ from $\chi_{cJ}$ since the experimental
analysis technique to separate the prompt signal is well developed already. 
The polarization parameter for $J/\psi$ from $\chi_{cJ}$ decay, which
is predicted in Fig. 1(b) as well as that for direct $J/\psi$ 
can also be measured.
Since CDF successfully separated the $\chi_{c1}$ and $\chi_{c2}$ signal
in their total cross section measurement,
the $p_T$ distributions of both $\chi_{c1}$ and $\chi_{c2}$ can be measured
separately.  The heavy-quark spin symmetry which has been used for the
simplification of $\chi_{cJ}$ ME's can be tested with higher accuracy.

If the polarization data continues to disagree with the predictions
of the NRQCD factorization approach, it would indicate a serious flaw in our
understanding of inclusive charmonium production.
If it is due to the fact that $m_c$ is too small, it might be interesting
to test the polarization of $\Upsilon$.  The recent measurement of the 
$\Upsilon$ production rate by CDF 
will allow for a better understanding of the NRQCD matrix elements for the
$\Upsilon$ system without many assumptions which were necessary due to
the lack of statistics. The theoretical prediction of the $\Upsilon$ 
polarization is needed both for Run I and Run II. We might also have
a chance to probe the gluon fragmentation dominance in high-$p_T$ 
$\Upsilon$ production in Run II even though the mass of $\Upsilon$ is
three times larger than that of $J/\psi$.

We expect that Run II of the Tevatron will definitely
produce significant information which will help us to understand
charmonium and bottomonium.

\section*{Acknowledgments}
The author thanks Eric Braaten and Bernd A. Kniehl for their enjoyable
collaboration on the subject discussed here, and reading the manuscript. 
This work was supported in part by the Alexander von Humboldt Foundation.



\section*{References}
\begin{thebibliography}{99}

\bibitem{B-B-L}
G.T. Bodwin, E. Braaten, and G.P. Lepage, \Journal{\PRD}{51}{1125}{1995};
{\bf 55}, 5855(E) (1997).

\bibitem{B-F}
E. Braaten and S. Fleming, \Journal{\PRL}{74}{3327}{1995}.

\bibitem{CDF-psi}
CDF Collaboration, \Journal{\PRL}{79}{572}{1997}.

\bibitem{B-Y}
E. Braaten and T. C. Yuan \Journal{\PRL}{71}{1673}{1993}.

\bibitem{Cho-Wise}
P. Cho and M.B. Wise, \Journal{\PLB}{346}{129}{1995}.



\bibitem{CDF-psipol} CDF Collaboration, hep-ex/0004027.

\bibitem{Beneke-Rothstein}
M. Beneke and I.Z. Rothstein, \Journal{\PLB}{372}{157}{1996};
        {\bf 389}, 769(E) (1996).

\bibitem{Beneke-Kramer}
M. Beneke and M. Kr\"amer, \Journal{\PRD}{55}{5269}{1997}.

\bibitem{Leibovich}
A. Leibovich, \Journal{\PRD}{56}{4412}{1997}.

\bibitem{CDF-chi} CDF Collaboration, \Journal{\PRL}{79}{578}{1997}.
\bibitem{CDF-chi12} CDF Collaboration, CDF Note 3121.

\bibitem{BKL}E. Braaten, B. Kniehl and J. Lee, hep-ph/9911436.

\bibitem{Cho-Leibovich}
P. Cho and A. Leibovich, \Journal{\PRD}{53}{150}{1996};
        {\it ibid.} {\bf 53}, 6203 (1996).
\bibitem{Maltoni}
F. Maltoni, private communication
on world average of $\Gamma(\chi_{c2}\to\gamma\gamma)$.


\bibitem{Kniehl-Kramer}
B.A. Kniehl and G. Kramer, {\it Eur. Phys. J.} {\bf C 6}, 493 (1999);
\Journal{\PRD}{60}{014006}{1996}.

\bibitem{PDF}
A.D. Martin, R.G. Roberts, W.J. Stirling, and R.S. Thorne,
{\it Eur. Phys. J.} {\bf C 4}, 463 (1998);
CTEQ Collaboration, hep-ph/9903282.

\bibitem{BL}
E. Braaten and J. Lee, hep-ph/0004228.

\end{thebibliography}
\end{document}